\documentclass[fleqn]{elsart5p}

\pdfoutput=1

\usepackage{graphicx}
\usepackage{latexsym}
\usepackage{amsbsy}
\usepackage{amsfonts}
\usepackage{amssymb}
\usepackage{amscd}
\usepackage{amsmath}
\usepackage{color}
\usepackage{rotating}

\usepackage{caption}
\usepackage{cases}
\usepackage{pst-all}

\usepackage[utf8]{inputenc} 
\usepackage[T1]{fontenc}

\def\cha{\kern .6em{\sqcup \kern -1.12em \sqcup}\kern .6em}

\def\I{{\rm i}}
\def\vec{\boldsymbol}
\def\q{\rm} 
\def\qvb{\vec} 
\def\its{\it}

\def\cal{\mathcal}

\def\CD{\hbox{
$\bigcirc {\kern-.75em { 
\hbox{$\buildrel > \over {\buildrel \over{}}$}}}$}}
\def\CG{\hbox{
$\bigcirc {\kern-.75em { 
\hbox{$\buildrel < \over {\buildrel \over{}}$}}}$}}

\mathchardef\Gamma="0100
\mathchardef\Delta="0101
\mathchardef\Theta="0102
\mathchardef\Lambda="0103
\mathchardef\Xi="0104
\mathchardef\Pi="0105
\mathchardef\Sigma="0106
\mathchardef\Upsilon="0107
\mathchardef\Phi="0108
\mathchardef\Psi="0109
\mathchardef\Omega="010A

\def\appendix{\par
  \setcounter{section}{0}%
  \def\thesection{Appendix \Alph{section}}}

\newtheorem{theorem}{Theorem}{\bf}{\its}
\newtheorem{proposition}{Proposition}[section]{\bf}{\its}
\newtheorem{remark}{Remark}[section]{\bf}{\rm}

\begin{document}
\begin{frontmatter}
 
 \null \vskip -3cm
  
 \title{\bf Synthesis of an arbitrary elliptical polarization operator  \\}
\author{Pierre Pellat-Finet}
\address{Universit\'e Bretagne Sud,  UMR  CNRS  6205, LMBA, F-56000 Vannes, France \\ pierre.pellat-finet@univ-ubs.fr}


 \begin{abstract} We prove a theorem for transforming the polarization eigenstates of an arbitrary elliptical birefringent device into the eigenstates of another similar device by means of a birefringent device.
   The theorem is applied to synthesize a specific birefringent device from a circular birefringent device, and a practical setup is described. The resulting birefringence and birefringent axis of the synthesized device can be independently set or modulated. Finally, the theorem is extended to  elliptical dichroic devices and elliptical polarizers.

  \begin{keyword}   Circular birefringence, elliptical birefringence, elliptical dichroism,  elliptical polarizer, equivalent rectilinear birefringence, polarization optics, quaternionic representation of polarized light, rotatory power.
\end{keyword}

\end{abstract}

\end{frontmatter}

\setcounter{footnote}{1}
\section{Introduction}\label{sect1}
By using appropriate phase plates, we can transform the rectilinear polarization eigenstates of a birefringent medium into elliptical polarization states, thereby converting the initial rectilinear birefringent device into an elliptical one  \cite{PPF1}. In this article, we will prove a theorem that solves the more general problem of transforming an arbitrary elliptical birefringent device into another one.
The result is applied to transforming a variable rotatory power into an elliptical birefringent device whose birefringence and eigenvibrations are independently adjustable.
We will also extend the previous theorem to elliptical polarizers and elliptical dichroic devives.

\section{Conventional definitions and results on polarized light}

\begin{figure}
  \begin{center}
  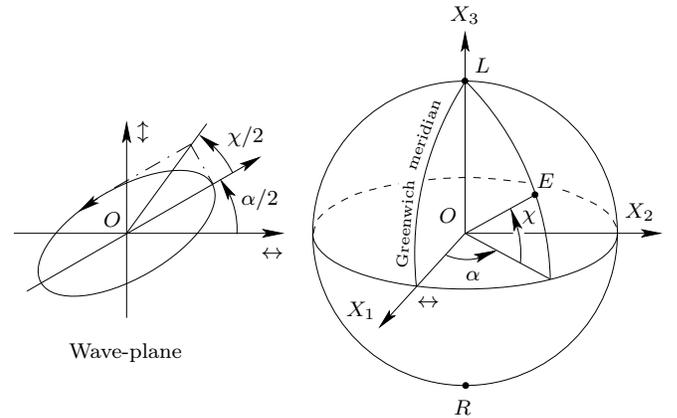
  \caption{Poincar\'e sphere representation of a polarization state. The point $E$ of latitude $\alpha$ and longitude $\chi$ represents the oriented-contour ellipse ${\mathsf E}(\alpha /2,\chi /2)$. The ellipse shown on the diagram is left-handed.\label{fig0}}
  \end{center}
\end{figure}

We consider harmonic polarized lightwaves propagating in a given direction in the physical space. This direction is assumed to be ``horizontal'', so that wave surfaces  are vertical planes,  orthogonal to the propagation direction. A polarization state is characterized by an oriented-contour ellipse  ${\mathsf E}$ whose major axis makes an angle $\alpha /2$ with the horizontal, in a wave-plane, and whose ellipicity (more precisely, ``angle of ellipticity'') is $\chi /2$, with $|\tan (\chi /2)|= b/a$, where $a$ and $b$  are the major and the minor axis lengths of the ellipse \cite{PPF1,Ram} (see Fig.\ \ref{fig0}). By convention we  choose $\chi >0$ for left-handed ellipses (left-handed vibrations). We denote the ellipse by ${\mathsf E} (\alpha /2,\chi /2)$.

We use the Poincar\'e sphere representation of polarized light \cite{Ram}, see Fig.\ \ref{fig0}. Every point on the sphere represents a unit polarization state (i.e. of unit irradiance). The sphere has two poles: the North Pole $L$ represents the (unit) left-circularly polarized state, and the South Pole $R$, the right-circularly polarized state. The equator is the set of rectilinearly polarized states. A ``Greenwich'' meridian is chosen as origin for longitudes and it intercepts the equator at a point representing the rectilinear state that is  horizontally polarized  in a wave-plane (and graphically indicated by  $\leftrightarrow$).  Longitudes are positive if taken from West to  East (unlike geographic conventions on the Earth). The point $E$, with longitude $\alpha$ and latitude $\chi$, corresponds to the ellipse ${\mathsf E}(\alpha /2,\chi /2)$.  We write $E(\alpha ,\chi )$ and say ``point $E$\,'' or, abusively, ``polarization state $E$.'' A state is elliptically polarized if $0<|\chi |<\pi /2$.

The ellipse ${\mathsf E}_\perp [(\alpha +\pi) /2,-\chi /2]$ is said to be orthogonal to the ellipse ${\mathsf E} (\alpha/2,\chi/2 )$. Points $E_\perp (\alpha +\pi,-\chi )$ and $E(\alpha ,\chi )$ are then opposite on the sphere, and the segment $E_\perp E$ is an axis of the sphere. The polarization states $E$ and $E_\perp$ are said to be orthogonal to each other.

A physical device that modifies the polarization state of an incident lightwave is called a polarization operator. (``Polarization operator'' thereby (abusively) designates both a physical device and a mathematical object.) The identity operator, denoted as ${\cal I}$, is considered as a polarization operator. If the output polarization state is identical to the input state, up to a multiplicative factor (called the eigenvalue), the state is an eigenstate of the operator.

An elliptical birefringent operator is a perfectly transparent  device  such that:
\begin{enumerate}
\item It admits two orthogonal unit polarization eigenstates $E$ and $E_\perp$ that are elliptically polarized.
\item It introduces a phase shift between the two eigenstates. This phase shift is called the birefringence.
\end{enumerate}

The phase shift has its origin in different phase velocities for waves propagating across the device, according to their polarization states. Points $E$ and $E_\perp$ define an axis $\Delta$ on the Poincaré sphere, called the operator axis. We choose to orientate the axis from the slow vibration to the fast one.

We have then \cite{Ram}:

\medskip
\begin{proposition} Let ${\cal B}$ be an elliptical  birefringent device with birefringence $\psi$ and axis $\Delta$ (on the Poincar\'e sphere). Let $P$ be the point on the Poincar\'e sphere representing  an arbitrary polarization state incident on ${\cal B}$. The emerging polarization state is then represented by the point $P'$\!,  obtained from $P$ by the rotation of angle $\psi$ around the axis $\Delta$. (The rotation follows the trigonometric direction---i.e. counterclockwise---for an observer to whom the axis is oriented.)
\end{proposition}

\medskip
Particular birefringent operators are:
\begin{itemize}
\item Rectilinear birefringent devices for which $E$ and $E_\perp$ belong to the equator. On the Poincar\'e sphere, they operate as rotations around equatorial axes.
\item Circular birefringent devices for which $E$ and $E_\perp$ are the poles of the Poincar\'e sphere. They operate as  rotations around the pole axis. A rectilinear state is rotated by an angle  equal to the birefringence $\psi$ and remains rectilinear. In a wave-plane, the effect is known as rotatory power. A rotatory power equal to $\rho$ corresponds to a circular birefringent device whose birefringence is $\psi =2\rho$. The term ``optical activity'' is used as equivalent to ``rotatory power.'' 
\end{itemize}

\smallskip
\begin{remark} {\rm If $\psi =0 \mod 2\pi$, the birefringent operator is the identity. In the following, considered operators are proper birefringent devices with $\psi \ne 0\mod 2\pi$.}
  \end{remark}

\section{Synthesizing a specific elliptical birefringent device}

\subsection{Changing-of-axis theorem: birefringent devices}

\begin{theorem}\label{th1} {Let ${\cal E}$ be an elliptical birefringent operator with axis $E_\perp E$ on the Poincar\'e sphere and elliptical birefringence $\psi$ ($\psi \ne 0\mod 2\pi$). Let ${\cal B}$ be a birefringent device that transforms $E$ into $E'$,
    and let ${\cal B}^{-1}$ be the inverse operator (such that ${\cal B}\circ{\cal B}^{-1}={\cal I}$). Then:
    \begin{enumerate}
    \item ${\cal B}$ transforms $E_\perp$ into the polarization state $E'_\perp$ that is orthogonal to $E'$ (i.e. point $E'_\perp$ is opposite to $E'$ on the Poincar\'e sphere).
    \item The product $\,{\cal E}'\!={\cal B}\circ{\cal E}\circ{\cal B}^{-1}$ is an elliptical bire\-frin\-gent operator with birefringence  $\psi$ and axis  $E'_\perp E'$.
  \end{enumerate}}
  \end{theorem}

\medskip
\noindent{\its Proof}. Since $E$ is transformed into $E'$, we write $E'={\cal B}(E)$, where $E$ and $E'$ are points on the Poincar\'e sphere (representing polarization states), and ${\cal B}$ is a rotation.

\smallskip
    \noindent {\its (i)} The Poincaré sphere is a subset of a 3-dimensional Euclidean vector space ${\mathbb E}_3$, and a rotation on ${\mathbb E}_3$ is a linear mapping.  Let $\vec{OE}$ denote the vector belonging to ${\mathbb E}_3$, with origin at $O$ and end at $E$. If $\vec{OE}$ is transformed into $\vec{OE'}$\!, then  $\vec{OE}_\perp =-\vec{OE}$ is transformed into $\vec{OE}'_\perp =-\vec{OE}'$\!, by linearity, and $E'_\perp$ is opposite to $E'$ on the Poincar\'e sphere. Polarization states $E'$ and $E'_\perp$ are orthogonal to each other.

\smallskip
\noindent {\its (ii)} Since ${\cal E}$ and ${\cal B}$ are rotations, the product ${\cal E}'$ is also a rotation.

Let us choose $E'$ as  representing the polarization state of a lightwave incident on ${\cal E}'$. The operator ${\cal B}^{-1}$ is a rotation on the Poincar\'e sphere, such that ${\cal B}^{-1} (E')=E$, and since $E$ belongs to  the axis of ${\cal E}$, we obtain
${\cal E}\circ{\cal B}^{-1}(E')=E$, and then ${\cal E}'(E')={\cal B}(E)=E'$. We conclude that $E'$ represents an eigenstate of ${\cal E}'$. The same reasoning shows that the point $E'_\perp={\cal B}(E_\perp)$  also represents an eigenstate. Then $E'_\perp E'$ is the axis of ${\cal E}'$\!.

Let $P$ be a point on the great circle $\Gamma$ whose plane is orthogonal to the axis $E_\perp E$, on the Poincar\'e sphere, and let $Q={\cal E}(P)$. The point $Q$ belongs to $\Gamma$ and $\angle POQ=\psi$.  Let $P'={\cal B}(P)$ and $Q'={\cal B}(Q)$. Since ${\cal B}$ is a rotation, both $P'$ and $Q'$ belong to $\Gamma '={\cal B}(\Gamma)$, the great circle whose plane is orthogonal to $E'_\perp E'$. We have $\angle P'OQ'=\psi$, which is the angle of the rotation ${\cal E}'$. The proof is complete.  \hfill $\qed$

\subsection{Analogy with a change of basis in linear algebra}

According to Theorem \ref{th1} the transformation law of polarization operators takes the form
\begin{equation}
  {\cal E}'={\cal B}\circ{\cal E}\circ{\cal B}^{-1}\,,\label{eq1}\end{equation}
and is related to the conventional matrix transformation under a change of basis as we show below.

We consider the 2-dimensional complex vector space of polarization states.
If ${\cal A}$ is a linear operator, we denote $[{\cal A}]$  its matrix  in the basis $\{{\qvb \varepsilon}_1,{\qvb \varepsilon}_2\}$, and $[{\cal A}]'$ its matrix in the basis $\{{\qvb \varepsilon}'_1,{\qvb \varepsilon}'_2\}$. They are such that \cite{Hal}
\begin{equation}
   [{\cal A}]'=C^{-1}\,[{\cal A}]\,C=C^{-1}\,[{\cal A}]\,C\,,\label{eq2}\end{equation}
where $C$ is the change-of-basis matrix from the first basis to the second one, that is, $C=(c_{ij})$, with
${\qvb \varepsilon}'_j=c_{1j}{\qvb \varepsilon}_1+c_{2j}{\qvb \varepsilon}_2$ and $|c_{11}c_{22}-c_{12}c_{21}|=1$. (The row index is $i$ and the column index is $j$.)

Let $[{\cal E}]$ denote the matrix of ${\cal E}$ in the basis $\{{\qvb \varepsilon}_1,{\qvb \varepsilon}_2\}$, and   $[{\cal E}]'$ in the basis   $\{{\qvb \varepsilon}'_1,{\qvb \varepsilon}'_2\}$. According to Eq.\ (\ref{eq2}) we have
\begin{equation}
  [{\cal E}]'=C^{-1}\,[{\cal E}]\,C\,.\label{eq3}\end{equation}

The problem addressed here is: to transform ${\cal E}$ into ${\cal E}'$, where  ${\cal E}'$ is with respect to $\{{\qvb \varepsilon}'_1,{\qvb \varepsilon}'_2\}$  as ${\cal E}$ with respect to $\{{\qvb \varepsilon}_1,{\qvb \varepsilon}_2\}$. We should have then
\begin{equation}
  [{\cal E'}]'= [{\cal E}]\,,\label{eq4}\end{equation}
where  $[{\cal E'}]'$ denotes the matrix of ${\cal E}'$ in the basis  $\{{\qvb \varepsilon}'_1,{\qvb \varepsilon}'_2\}$. 

We apply Eq.\  (\ref{eq3}) to ${\cal E}'$ (we replace ${\cal E}$ with ${\cal E}'$), and according to Eq.\ (\ref{eq4}), we obtain
\begin{equation}
 C^{-1}\,[{\cal E}']\,C= [{\cal E}]\,,
\end{equation}
that is,
\begin{equation}
[{\cal E}']=C \,[{\cal E}]\,C^{-1}\,.\label{eq6}
\end{equation}
(Remark: we have to differentiate between  the matrix of ${\cal E}'$ in the basis $\{{\qvb \varepsilon}_1,{\qvb \varepsilon}_2\}$, denoted  as $[{\cal E}']$, and the matrix of ${\cal E}$ in the basis  $\{{\qvb \varepsilon}'_1,{\qvb \varepsilon}'_2\}$, denoted as $[{\cal E}]'$, so that Eqs.\ (\ref{eq3})  and (\ref{eq6}) not contradictory.)

Let us denote ${\cal B}$ the change-of-basis operator from   $\{{\qvb \varepsilon}_1,{\qvb \varepsilon}_2\}$ to  $\{{\qvb \varepsilon}'_1,{\qvb \varepsilon}'_2\}$. Its matrix in the basis $\{{\qvb \varepsilon}_1,{\qvb \varepsilon}_2\}$ is $C$, as previously defined, that is, $[{\cal B}]=C$.  Equation (\ref{eq6}) becomes
\begin{equation}
  [{\cal E}']=[{\cal B}] \,[{\cal E}]\,[{\cal B}]^{-1}\,,\label{eq7}\end{equation}
which is the matrix form of Eq.\ (\ref{eq1}) (in the basis $\{{\qvb \varepsilon}_1,{\qvb \varepsilon}_2\}$).

\smallskip
\begin{remark} {\rm Matrices $[{\cal E}']$ and $[{\cal E}]$ in Eq.\ (\ref{eq7}) refer to the basis  $\{{\qvb \varepsilon}_1,{\qvb \varepsilon}_2\}$, whereas in Eq.\ (\ref{eq3}), $[{\cal E}]'$ refers to $\{{\qvb \varepsilon}'_1,{\qvb \varepsilon}'_2\}$, and $[{\cal E}]$  to $\{{\qvb \varepsilon}_1,{\qvb \varepsilon}_2\}$.}
  \end{remark}
    
\smallskip
\begin{remark} {\rm To some extend, this section may be seen as a proof of Theorem \ref{th1}.}
  \end{remark}

\subsection{Application to synthesizing an elliptical birefringent}\label{sect33}

We apply the previous theorem to convert a rotatory power into a predefined elliptical birefringent device of birefringence $\psi$ and axis $\Delta$. On the Poincar\'e sphere, the latitude of $\Delta$ is $\chi$ and its longitude is $\alpha$.  The elliptical birefringent device is thus caracterized by 3 independent parameters: $\alpha$, $\chi$ and $\psi$.

\begin{figure}[b]
  \begin{center}
  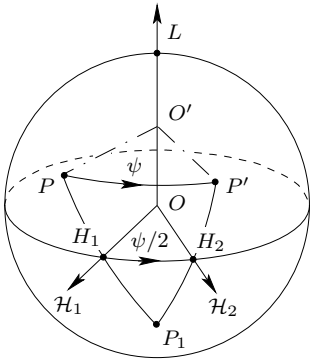
  \caption{The product of two half-wave plates ${\cal H}_1$ and ${\cal H}_2$ is equi\-va\-lent to a rotatory power. If $\angle H_1OH_2=\psi /2$,  the state $P$ is then transformed into $P'$ with $\angle PO'P'=\psi$. In a wave-plane, the fast axes of the two half-wave plates are oriented at $\psi /4$ to each other.
\label{fig1}}
  \end{center}
\end{figure}

It is well known that the product of two half-wave plates is equivalent to a rotatory power. The rotatory power is equal to $\rho$, if  the angle  between the fast vibrations of the two wave plates is $\rho /2$ (in a wave-plane).
To obtain then a circular birefringence $\psi$, we use two half-wave plates, say ${\cal H}_1$ and ${\cal H}_2$, whose axes on the Poincar\'e sphere makes an angle $\psi /2$. We have (Fig. \ref{fig1})
\begin{equation}
P\stackrel{{\cal H}_1}{\longrightarrow}P_1\stackrel{{\cal H}_2}{\longrightarrow}P'.
\end{equation}

In the physical space, the fast axes of ${\cal H}_1$ and ${\cal H}_2$ make an angle $\psi /4$ between them. Only the relative orientation of the two wave plates matters. If the rapid axis of ${\cal H}_1$ is along $\eta$ (in the physical space), the axis of ${\cal H}_2$ is obtained from $\eta$ by the rotation of angle $\psi /4$, in the trigonometric direction (corresponding to positive $\psi$,  see Fig.~\ref{fig3}). A tunable optical activity is obtained by changing the orientation of one of the two wave plates.

The product of ${\cal H}_1$ and ${\cal H}_2$ is  a circular birefringent device of birefringence $\psi$ (or a rotatory power $\psi /2$), and we write
\begin{equation}
  {\cal E}={\cal H}_2\circ{\cal H}_1\,.\end{equation}

\begin{figure}[b]
  \begin{center}
  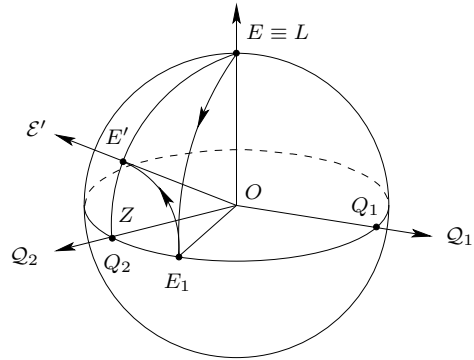
  \caption{Transforming the left-circular polarization state (North Pole $L$) into a given elliptical state $E'$ (latitude $\chi$) with two quarter-wave plates ${\cal Q}_1$ and ${\cal Q}_2$. We have $\angle Q_2OE_1=\angle Q_2OE'=\chi$, and $\angle E_1OQ_1=\pi /2$.\label{fig2}}
  \end{center}
\end{figure}

\begin{figure}[t]
  \begin{center}
  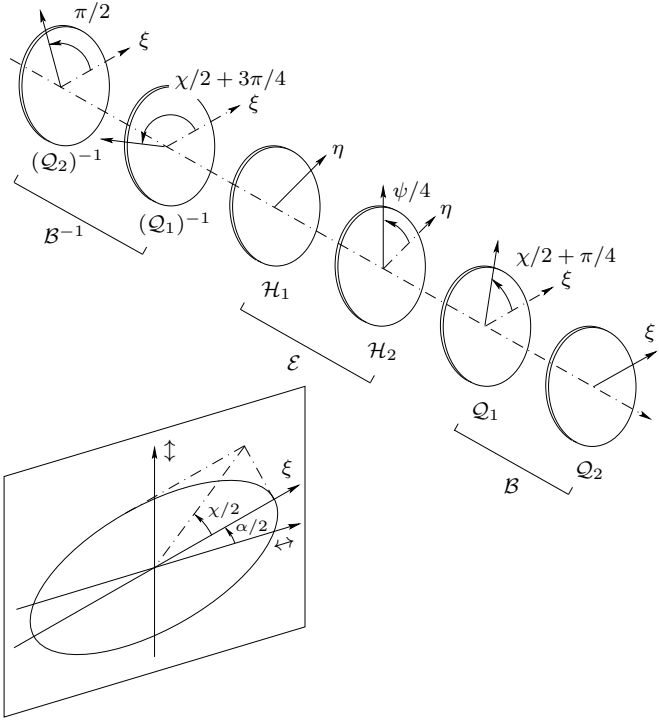
  \caption{Transforming a rotatory power into an elliptical birefringent device (birefringence $\psi$). Left bottom: oriented-contour ellipse representing the fast vibration of the desired elliptical birefringent in a wave-plane. The ellipse major axis is $\xi$ ; the ellipticity is $\chi /2$.  The rotatory power,  made up of two half-wave plates ${\cal H}_1$ and ${\cal H}_2$, is transformed into an elliptical birefringent device by means of four quarter-wave plates. The fast axis of every wave plate is shown. The angle between fast axes of ${\cal H}_1$ and ${\cal H}_2$ is equal to  $\psi /4$.  The fast axis of ${\cal Q}_2$ is aligned with the major axis $\xi$ of the ellipse. The orientations of  fast axes of the other wave plates are indicated with respect to $\xi$.  Light propagates from top-left to bottom-right.\label{fig3}}
  \end{center}
\end{figure}

To obtain an elliptical birefringent operator with axis $\Delta$ and birefringence 
$\psi$, we have to transform the North Pole of the sphere into one of the two eigenstates of the birefringent device. This can be achieved using two quarter-wave plates ${\cal Q}_1$ and ${\cal Q}_2$ as shown in Fig.\ \ref{fig2} and as explained below. (Wave plates ${\cal Q}_1$ and ${\cal Q}_2$ are rectilinear birefringent devices.)

Let $E'$ be the fast eigenstate of the elliptical device to be synthesized, let $\xi$ be the major axis of the corresponding oriented-contour ellipse (in  a wave-plane, see Fig. \ref{fig3}) and let  $\chi /2$ be its ellipticity. On the Poincaré sphere, the axis $\xi$ is represented by the equatorial axis $OZ$, which  intercepts the equator at $Z$  (Fig. \ref{fig2}), with longitude $\angle X_1OZ=\alpha$.  Then $E'$ belongs to the great circle passing by $Z$ and $L$ and its latitude is $\chi$ (that is, $\angle ZOE'=\chi$).  The wave plate ${\cal Q}_2$ is oriented so that the representative point  of its fast vibration, on the Poincar\'e sphere, is $Q_2$, with $Q_2\equiv Z$. (In practice this means that the quarter-wave plate ${\cal Q}_2$ is aligned with the major axis $\xi$ of the previous ellipse.) 
Let $E_1$ belong to the equator, with $\angle Q_2OE_1=\chi$, and let $Q_1$, the representative point of the fast vibration of ${\cal Q}_1$, be such that  $\angle E_1OQ_1=\pi /2$. The North Pole $L$ is transformed into $E'$ under the product (in the order) of the rotation of angle $\pi/2$ around $OQ_1$  and the rotation of angle $\pi /2$ around $OQ_2$. 
We have
\begin{equation}
  L\stackrel{{\cal Q}_1}{\longrightarrow}E_1\stackrel{{\cal Q}_2}{\longrightarrow}E',
\end{equation}
which means that the operator  ${\cal B}={\cal Q}_2\circ{\cal Q}_1$ transforms $L$ into $E'$, that is, the fast eigenstate of the circular birefringent device ${\cal E}$ into the fast eigenstate of the elliptical birefringent device to be synthesized.

The inverse operator of ${\cal B}$ is ${\cal B}^{-1}=({\cal Q}_1)^{-1}\circ ({\cal Q}_2)^{-1}$\!, where $({\cal Q}_j)^{-1}$ is the inverse operator of ${\cal Q}_j$ ($j=1,2$). Since ${\cal Q}_j$ is a quarter-wave plate, $({\cal Q}_j)^{-1}$ is a quarter-wave plate whose fast axis is at $\pi /2$ (or equivalently at $-\pi /2$) from the fast axis of ${\cal Q}_j$, in a wave-plane (see Fig.\ \ref{fig3}). On the Poincar\'e sphere, the axes of ${\cal Q}_j$ and $({\cal Q}_j)^{-1}$ are collinear with opposite orientations.

 According to Theorem \ref{th1} ve have
\begin{eqnarray}
  {\cal E}'&=&{\cal B}\circ{\cal E}\circ{\cal B}^{-1}  \nonumber \\
  &=&{\cal Q}_2\circ{\cal Q}_1\circ{\cal E}\circ
  ({\cal Q}_1)^{-1}\circ({\cal Q}_2)^{-1} \nonumber \\
  &=&{\cal Q}_2\circ{\cal Q}_1\circ{\cal H}_2\circ{\cal H}_1\circ
  ({\cal Q}_1)^{-1}\circ({\cal Q}_2)^{-1}\,.
  \end{eqnarray}

The synthesized elliptical birefringent device (birefringence $\psi$,  axis with longitude $\alpha$ and latitude $\chi$)  is shown in Fig.\ \ref{fig3}.

\subsection{Modulated elliptical birefringent device}

In the previous device, the elliptical birefringence $\psi$ and the birefringent axis $\Delta$ are independently synthesized. The birefringence $\psi$ only depends on the relative orientations of wave plates ${\cal H}_1$ and ${\cal H}_2$, and the axis $\Delta$ only depends on ${\cal Q}_1$ and $\cal Q_2$.  Then:
\begin{enumerate}
\item Given the axis $\Delta$ (i.e. given $\alpha$ and $\chi$), a variation of $\psi$ is obtained by varying the orientation of ${\cal H}_2$ (or equivalently that of ${\cal H}_1$).
\item Given the birefringence $\psi$ :
  \begin{enumerate}
    \item A variation of the latitude $\chi$ of $\Delta$ (without changing its longitude $\alpha$) is obtained by varying the orientation of ${\cal Q}_1$ (and that of $({\cal Q}_1)^{-1}$ accordingly).
    \item A variation of the longitude $\alpha$ of $\Delta$ (without changing its latitude $\chi$) is obtained by rotating ${\cal Q}_1$ and ${\cal Q}_2$ at the same time and maintaining constant their relative orientation. (Wave plates  $({\cal Q}_1)^{-1}$ and  $({\cal Q}_2)^{-1}$ are rotated accordingly.)
      \end{enumerate}
\end{enumerate}

Parameters $\psi$, $\chi$ and $\alpha$ can be tuned independently of each other. This is an important property in the context of obtaining a tunable birefringent device with the possibility of modulating either the phase or the axis. (Modulating the axis means modulating either its latitude or its longitude.)

We provide two examples:
\begin{enumerate}
\item Let us assume that the wave-plate ${\cal H}_2$ is counterclockwise continuously rotated, so that its angle with the axis $\eta$ is $f (t)=\omega t$, where $t$ denotes the time and $\omega$ the angular rotation velocity.  The other wave-plates are fixed. The resulting elliptical birefringent device has then a fixed axis (on the Poincar\'e sphere) and a varying birefringence equal to $\psi (t)=4f(t)=4\omega t$.
\item  Let us assume that the wave-plate ${\cal Q}_1$ is counterclockwise continuously rotated, its angle with the axis $\xi$ being $g(t)=\omega t +(\pi /4)$, and that  $({\cal Q}_1)^{-1}$ is also  counterclockwise continuously rotated, its angle with the axis $\xi$ being $h(t)=\omega t +(3\pi /4)$. Other wave-plates are fixed. The resulting elliptical birefringence is fixed, whereas the birefringent axis describes the meridian circle passing by $Q_2$ (see Fig.\ \ref{fig2}), its latitude being $\chi (t)=2\omega t$.
  \end{enumerate}

\section{A more general theorem}

We now extend Theorem \ref{th1} to polarizers and dichroic devices, and, more generally, to complex-phase shifters. The statement of the new theorem, as well as its proof, will refer to the quaternionic representation of polarized light.

\subsection{A brief overview of the quaternionic representation of polarized light}

A polarization state is represented by a quaternion of the form \cite{PPF1,PPF2,PPF3}
\begin{equation}
  X=X_0{\q e}_0+ \I X_1{\q e}_1+ \I X_2{\q e}_2+ \I X_3{\q e}_3\,,
\end{equation}
where the $X_\mu$'s are the Stokes parameters of the considered polarized lightwave \cite{Ram} and where $\{ {\q e}_0,{\q e}_1,{\q e}_2,{\q e}_3\}$ is the canonical basis of the complex quaternion algebra.  Quaternions of the form $X_0{\q e}_0$ can be seen as real numbers, that is ${\q e}_0\equiv 1$. Unit quaternions ${\q e}_1$, ${\q e}_2$ and ${\q e}_3$ can be seen as unit vectors along axes $X_1$, $X_2$ and $X_3$ of the Poincar\'e sphere (see Fig.\ \ref{fig0}). The vector $\vec{OE}$ in Fig.\ \ref{fig0} can then be identified with the unit quaternion ${\q e}_n$ such that
\begin{equation}
  {\q e}_n={\q e}_1\cos\chi\cos\alpha + {\q e}_2\cos\chi\sin\alpha + {\q e}_3\sin\chi\,,\end{equation}
and the polarization state represented by $E$ on the Poincar\'e sphere is represented by the quaternion ${\q e}_0+\I{\q e}_n$. The (unit) orthogonal state $E_\perp$ is represented by ${\q e}_0 -\I{\q e}_n$.  The axis $E_\perp E$ on the Poincar\'e sphere is collinear to ${\q e}_n$, and we also say ``axis ${\q e}_n$.''

A pure phase-shifter (or pure ``dephasor'' \cite{PPF2}), whose eigenstates are $E$ and $E_\perp$, 
is represented by
\begin{equation}u=\exp {{\q e}_n \psi \over 2}={\q e}_0\cos{\psi\over 2}+{\q e}_n\sin{\psi\over 2}\,,
  \end{equation}
where ${\q e}_n$ is the axis of the operator and $\psi$ is the complex phase-shift. We have $\psi =\varphi +\I \delta$, where $\varphi$ is the birefringence and $\delta$ the dichroism. If $\delta =0$, the operator is a pure birefringent operator, and if moreover $\varphi >0$, the state $E$ (or ${\q e}_0+\I{\q e}_n$) corresponds to  the fast eigenstate. 

A polarizer is represented by a quaternion of the form
\begin{equation}
  u={1\over 2}({\q e}_0+\I{\q e}_n)\,,\end{equation}
where ${\q e}_n$ represents the transmitted state, on the Poincar\'e sphere. 

Physical polarization operators generally exhibit additional isotropic absorption. If $\kappa$ ($\kappa >0$) denotes the isotropic  absorption factor, then a physical polarizer is represented by
\begin{equation}
  u={\sqrt{\kappa}\over 2}({\q e}_0+\I{\q e}_n)\,,\end{equation}
and a physical phase-shifter by
\begin{equation}
  u=\sqrt{\kappa} \,\exp {{\q e}_n \psi \over 2}\,.\end{equation}
For a physical dichroic device we have  $\psi =\I\delta$ ($\varphi =0$) with  $\sqrt{\kappa} \,\exp(\pm \delta /2) <1$. For $\delta >0$, the  less absorbed polarization state is $E$, represented by ${\q e}_0+\I{\q e}_n$.

Polarization operators operate according to
\begin{equation}
  X'=u\,X\,\overline{u}^{\, *}\,,\label{eq19}\end{equation}
where  $\overline{u}^{\, *}$ is the complex conjugate of the Hamilton conjugate of $u$. (The quaternion $X$ represents the input state and $X'$ the output state.)

\subsection{General theorem}
\begin{theorem}\label{th2} Let ${\cal E}$ be a polarization operator with axis $E_\perp E$ on the Poincat\'e sphere. Let ${\cal B}$ be an elliptical birefringent operator that transforms $E$ into $E'$. Then:
  \begin{enumerate}
  \item ${\cal B}$ transforms $E_\perp$ (which is orthogonal to $E$) into the polarization state $E'_\perp$ orthogonal to  $E'$.
  \item${\cal E}'={\cal B}\circ{\cal E}\circ{\cal B}^{-1}$ is a polarization operator  with axis $E'_\perp E'$ and similar to ${\cal E}$; more precisely:
    \begin{enumerate}
    \item If ${\cal E}$ is a polarizer with isotropic absorbtion $\kappa$, then ${\cal E}'$ is a polarizer with isotopic absorption $\kappa$.
      \item If ${\cal E}$ is a phase shifter with complex phase-shift $\psi$ and isotropic absorption $\kappa$, then ${\cal E}'$ also is  a phase shifter with complex phase-shift $\psi$ and isotropic absorption $\kappa$.
      \end{enumerate}
    \end{enumerate}
\end{theorem}

\smallskip
\noindent{\its Proof}.   Let ${\cal B}$ be an elliptical birefringent operator that transforms $E$ into $E'$. In the quaternionic representation of polarized light, ${\cal B}$ is represented by a unit quaternion of the form
$w=\exp ( {\q e}_w\varphi /2)$, where ${\q e}_w$ is a real unit pure quaternion and $\varphi$ a real number. Then $\overline{w}^{\,*}=\exp (- {\q e}_w\varphi /2)=w^*$\!, and $ww^*=1={\q e}_0$. Let $X={\q e}_0+\I {\q e}_n$ be the quaternionic representation of $E$, and let
$X'={\q e}_0+\I {\q e}_m$ be that of $E'$. Since $E'={\cal B}(E)$,  we set $u=w$ in Eq. (\ref{eq19}) and we obtain (we use $\overline{w}^{\, *}=w^*$)
\begin{equation}
  {\q e}_0+\I {\q e}_m= w ({\q e}_0+\I {\q e}_n)w^*={\q e}_0+\I w {\q e}_nw^*,
\end{equation}
and then  ${\q e}_m= w{\q e}_n w^*$.

\smallskip
\noindent {\its (i)} 
The polarization state $E_\perp$ is represented by ${\q e}_0-\I {\q e}_n$ and is transformed into
\begin{equation}
  w ({\q e}_0-\I {\q e}_n) w^*= {\q e}_0-\I w{\q e}_n w^*={\q e}_0-\I {\q e}_m\,,\end{equation}
which represents the polarization state orthogonal to $E'$.

\smallskip
\noindent {\its (ii)}
If ${\cal E}$ is an elliptical  phase-shifter, with axis ${\q e}_n$, it  is represented by a quaternion of the form
$\sqrt{\kappa} \exp ({\q e}_n\psi /2)$, where $\psi$ is a complex number. The operator ${\cal E}'={\cal B}\circ{\cal E}\circ{\cal B}^{-1}$ is then represented by
\begin{eqnarray}
  u'&=&\exp\left({\q e}_w{\varphi\over 2}\right)\,\sqrt{\kappa}\,\exp\left({\q e}_n{\psi\over 2}\right) \exp\left(-{\q e}_w{\varphi\over 2}\right)\nonumber \\
  &=& \sqrt{\kappa}\,w\left({\q e}_0\cos{\psi\over 2}+{\q e}_n\sin{\psi\over 2}\right)w^*\nonumber \\
  &=&  \sqrt{\kappa}\,\left({\q e}_0\cos{\psi\over 2}+w{\q e}_nw^*\sin{\psi\over 2}\right)\nonumber \\
  &=&\sqrt{\kappa}\,\exp\left({\q e}_m{\psi\over 2}\right)\,.
\end{eqnarray}
And the theorem is proved in that case.

If ${\cal E}$ is an elliptical polarizer, whose axis is ${\q e}_n$ (on the Poincar\'e sphere), it is represented by a quaternion of the form  $u=(\sqrt{\kappa}/2)({\q e}_0+\I{\q e}_n)$. Then ${\cal E}'={\cal B}\circ{\cal E}\circ{\cal B}^{-1}$ is represented by
\begin{eqnarray}
  u'&=&{\sqrt{\kappa}\over 2}\exp\left({\q e}_w{\varphi\over 2}\right)({\q e}_0+\I{\q e}_n)\exp\left(-{\q e}_w{\varphi\over 2}\right)\nonumber \\
  &=& {\sqrt{\kappa}\over 2}({\q e}_0+\I{\q e}_m)\,.
\end{eqnarray}
The proof is complete. \hfill
$\qed$

\section{Another synthesis of an elliptical birefringent device}

An elliptical birefringent device ${\cal E}$ can be thought of as the product ${\cal E}={\cal B}' \circ{\cal R}'$ of a rotatory power ${\cal R}'$ and a rectilinear birefringent device ${\cal B}'$\!, called the equivalent rotatory power and the equivalent  rectilinear birefringence \cite{PPF1,PPF2}.

Let $\psi$ be the elliptical birefringence of ${\cal E}$ and let $\Delta$ be its axis, with longitude $\alpha$ and latitude $\chi$.   
Let $\theta '$ be the equivalent circular birefringence (equal to twice the equivalent rotatory power). Let $\varphi '$ be the equivalent rectilinear birefringence and let $\xi$ be the direction of its fast vibration in the physical space. The axis $\xi$ is represented  on the Poincar\'e sphere by the axis $OZ'$, with  $Z'$ belonging to the equator (longitude  $\alpha '$).  Those parameters are connected as follows \cite{PPF1,PPF2} (see Appendix for a proof)
\begin{equation}
  \sin{\varphi '\over 2}=\cos\chi\sin{\psi\over 2}\,,\label{eq14}\end{equation}
\begin{equation}
  \tan{\theta '\over 2}=\sin\chi\tan{\psi\over 2}\,,\label{eq15}\end{equation}
\begin{equation}
  \alpha '-{\theta '\over 2}=\alpha\,.\label{eq16}\end{equation}

The elliptical birefringent device ${\cal E}$ can be synthesized by separately synthesizing ${\cal B}'$ and ${\cal R}'$, as proposed by Pab\'on {\its et al.} \cite{Pab}. The equivalent birefringent device ${\cal B}'$ can be synthesized from a rotatory power  equal to $\varphi '/2$ (circular birefringence $\varphi '$) obtained as the product of two half-wave plates ${\cal H}_1$ and ${\cal H}_2$, whose fast axes make an angle $\varphi '/4$ in the physical space. For the axis $OZ'$, we apply the method of Section \ref{sect33} by setting $E'=Z'$. Since $Z'$ belongs to the equator, the point $E'=Z'$ merges with $E_1$ (see Fig.\ \ref{fig2}): $Z'=E'= E_1$,  so that ${\cal Q}_2$ is useless (Fig.\ \ref{fig2}) and is removed.
The quarter-wave plate ${\cal Q}_1$  transforms the circular state $L$ into the point $E_1= Z'$ corresponding to the fast vibration of ${\cal B}'$.

The equivalent rotatory power ${\cal R}'$ is obtained as the pro\-duct of two half-wave plates ${\cal H}_3$ and ${\cal H}_4$, with fast vibrations making an angle $\theta '/4$ in the physical space.

\begin{figure}[t]
  \begin{center}
  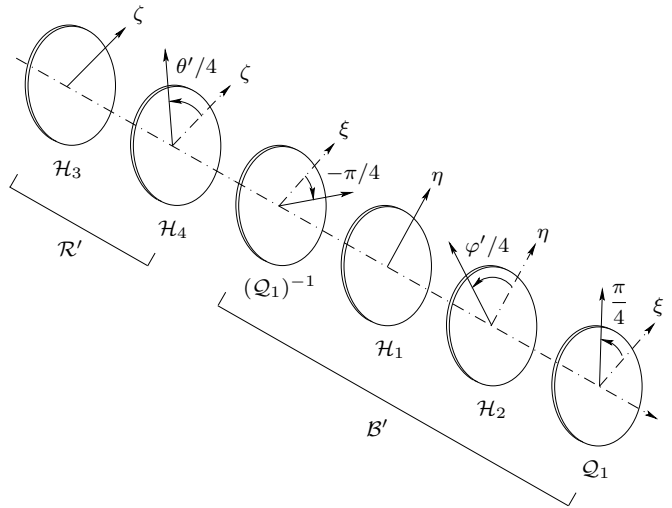
  \caption{Synthesis of an elliptical birefringent device through its equivalent optical activity ${\cal R}'$ and equivalent rectilinear birefringence ${\cal B}'$, as proposed by Pab\'on {\its et al.} \cite{Pab}.  Wave plates ${\cal H}_j$ are half-wave plates and ${\cal Q}_1$ is a quarter-wave plate. The wave plate $({\cal Q}_1)^{-1}$ is a quarter-wave plate, inverse to ${\cal Q}_1$: its fast axis is at $\pm \pi /2$ from the fast axis of ${\cal Q}_1$ (in the physical space). For each wave plate we indicate the orientation of the fast vibration by a  solid-line arrow. Direction $\eta$ and $\zeta$ are arbitrary. The axis $\xi$ corresponds to the orientation of the fast rectilinear eigenstate of ${\cal B}'$. Light propagates from top-left to bottom-right.\label{fig4}}
  \end{center}
\end{figure}

Figure \ref{fig4} shows the device as proposed by Pab\'on {\its et al.} \cite{Pab}.
The equivalent rotatory power ${\cal R}'$ and the equivalent birefringent device ${\cal B}'$ are produced independently: ${\cal R}'$ is obtained by adjusting the angle between ${\cal H}_3$ and ${\cal H}_4$; the equivalent birefringence $\varphi '$ is obtained by adjusting the angle between ${\cal H}_1$ and ${\cal H}_2$ ; the axis by setting  the orientation of ${\cal Q}_1$ and $({\cal Q}_1)^{-1}$ at $\pm \pi /4$ with respect to the desired birefringent axis (axis $\xi$ in Fig.\ \ref{fig4}). But the parameters of the initial elliptical birefringent device to be be synthesized (that is, $\psi$, $\chi$ and $\alpha $) are not obtained independenty of each other as in Section \ref{sect33}. This can be understood as follows.

Equations (\ref{eq14})--(\ref{eq16}) can be inverted in the form \cite{PPF1,PPF2}
\begin{equation}
  \cos{\psi\over 2}=\cos{\varphi '\over 2}\cos{\theta '\over 2}\,,\label{eq27}\end{equation}
\begin{equation}
  \tan \chi=\cot{\varphi '\over 2}\sin{\theta '\over 2}\,,\label{eq28}\end{equation}
\begin{equation}
  \alpha =\alpha '-{\theta '\over 2}\,,\label{eq29}\end{equation}
as shown in the Appendix.
Then $\psi$ and $\chi$ depend on both $\varphi '$ and $\theta '$. If we modify $\varphi'$ only, we modify both $\psi$ and $\chi$;  if we modify $\theta '$ only we also modify both $\psi$ and $\chi$. We cannot vary $\psi$ by adjusting only ${\cal R}'$ or only ${\cal B}'$: we have to adjust both $\varphi '$ and $\theta '$, and  $\chi$  is  also necessarily modified (in general).

Another drawback of the setup of Fig.\ \ref{fig4}, with respect to that of Fig.\ \ref{fig3}, is that the orientations of the wave-plates are not directly related to the parameters of the desired elliptical birefringent device. In Fig.\ \ref{fig3}, we remark that the orientation of each wave plate  is directly connected to only one of the parameters $\psi$, $\chi$ or $\xi$ (i.e. $\alpha$). In Fig.\ \ref{fig4}, the orientations depend on both $\psi$ and $\chi$ through $\varphi '$ and $\theta '$, which have to be calculated before appropriately adjusting each wave plate.

\section{Conclusion}
Theorems 1 and 2 hold, because the elliptical birefringent device ${\cal B}$ is no more than a rotation on the Poincar\'e sphere, which preserves orthogonality of polarization states. Another problem 
would be to transform an elliptical phase-shifter into an operator whose eigenstates are not orthogonal. This could be achieved with a dichroic device.  

\section*{Appendix}
To make the article self contained---as far as possible---we prove Eqs.\ (\ref{eq14})--(\ref{eq16}).

The elliptical birefringent device ${\cal E}$ is represented by $\exp ({\q e}_n\psi /2)$ and is the product ${\cal B}'\circ{\cal R}'$ of  the equivalent circular birefringence ${\cal R}'$, represented by
$\exp ({\q e}_3\theta ' /2)$, followed by  the equivalent rectilinear birefringent device ${\cal B }'$, represented by by
$\exp ({\q e}_m\varphi ' /2)$.
The axes of ${\cal E}$ and ${\cal B}'$  respectively are  ${\q e}_n$ and ${\q e}_m$ with
\begin{equation}
  {\q e}_n={\q e}_1\cos\chi \cos\alpha +{\q e}_2\cos\chi \sin\alpha+ {\q e}_3\sin\chi\,,\label{eq30}\end{equation}
and \begin{equation}
  {\q e}_m={\q e}_1\cos\alpha '+{\q e}_2\sin\alpha '\,.\end{equation}

We assume that $\psi$, $\chi$ and $\alpha$ belongs to $[-\pi, \pi]$ and will look for $\alpha '$, $\varphi '$ and $\theta '$ also in  $[-\pi, \pi]$.

We first remark that if $\chi=0\mod \pi$, the operator ${\cal E}$ is a rectilinear birefringent device: we choose ${\cal R}'={\cal I}$, so that ${\cal B}'\equiv{\cal E}$. The values of the equivalent parameters are   $\theta '=0$, $\alpha '=\alpha$ and $\varphi '=\psi$ : Eqs.\ (\ref{eq14})--(\ref{eq16}) hold true.

If $\chi =\pm\pi /2$, then ${\cal E}$ is a rotatory power, and we choose ${\cal B}'=I$ and ${\cal R}'\equiv{\cal E}$. We have $\theta '=\pm \psi$, $\varphi ' =0$, and $\alpha '$ is arbitrary since $\alpha$ is arbitrary. Eqs.\ (\ref{eq14})--(\ref{eq16}) hold true.

In the following, we will assume $\chi\ne 0\mod \pi /2$, and $\psi\ne 0\mod 2\pi$.

We use ${\q e}_m{\q e}_3=-{\q e}_m\vec \cdot{\q e}_3+{\q e}_m\vec \times {\q e}_3= {\q e}_m\vec \times {\q e}_3$ (because  ${\q e}_m\vec \cdot{\q e}_3=0$), and obtain 
\begin{eqnarray}
  \exp{{\q e}_n\psi\over 2}&=&\exp {{\q e}_m\varphi '\over 2}\exp {{\q e}_3\theta '\over 2}\nonumber \\
  &=& \left({\q e}_0\cos{\varphi '\over 2}+{\q e}_m\sin{\varphi '\over 2}\right)
  \left({\q e}_0\cos{\theta '\over 2}\right. \nonumber \\
  & & \hskip 4.2cm \left. +\,{\q e}_3\sin{\theta '\over 2}\right)\nonumber \\
  &=& {\q e}_0\cos{\varphi '\over 2}\cos{\theta '\over 2}
  +{\q e}_m\sin{\varphi '\over 2}\cos{\theta '\over 2}  \nonumber \\
  & &\hskip 2.35cm + \,{\q e}_m\times {\q e}_3 \sin{\varphi '\over 2} \sin{\theta '\over 2}
  \nonumber \\
  & &\hskip 2.35cm + \,{\q e}_3 \cos{\varphi '\over 2} \sin{\theta '\over 2}\,.\label{eq32}
  \end{eqnarray}
From $ \exp ({\q e}_n\psi/ 2)=\cos (\psi /2)+{\q e}_n\sin (\psi /2)$ and from Eq.\ (\ref{eq30}), and comparing the components on ${\q e}_0$ and ${\q e}_3$ in Eq.\ (\ref{eq32}),  we deduce
\begin{equation}
  \cos{\varphi '\over 2}\cos{\theta '\over 2}=\cos{\psi\over 2}\,,\label{eq33}
\end{equation}
and
\begin{equation}
  \cos{\varphi '\over 2}\sin{\theta '\over 2}=\sin\chi\sin{\psi\over 2}\,,\label{eq34}
\end{equation}
and since $\chi \ne 0$ and $\chi\ne \pm\pi$, we obtain 
\begin{equation}
  \tan{\theta '\over 2}=\sin\chi\tan{\psi\over 2}\,,\end{equation}
which is Eq.\ (\ref{eq15}). We remark that $\theta '$ is defined modulo $2\pi$.

From
\begin{equation}
  {\q e}_m\vec\times{\q e}_3={\q e}_1\sin\alpha '-{\q e}_2\cos\alpha '\,,
  \end{equation}
and comparing components on ${\q e}_1$, we deduce
\begin{eqnarray}
 \cos\alpha '\sin{\varphi '\over 2}\cos{\theta '\over 2}&+&
  \sin\alpha '\sin{\varphi '\over 2}\sin{\theta '\over 2}\nonumber \\
  & & \hskip .8cm = \cos\chi\cos\alpha \sin{\psi\over 2}\,,
\end{eqnarray}
that is
\begin{equation}
  \sin{\varphi '\over 2}\cos\left(\alpha '-{\theta'\over 2}\right)=
\cos\chi\cos\alpha \sin{\psi\over 2}\,.\label{eq37}
\end{equation}
Comparing components on ${\q e}_2$, we obtain
\begin{eqnarray}
 \sin\alpha '\sin{\varphi '\over 2}\cos{\theta '\over 2}&-&
  \cos\alpha '\sin{\varphi '\over 2}\sin{\theta '\over 2}\nonumber \\
  & & \hskip .8cm  =\cos\chi\sin\alpha \sin{\psi\over 2}\,,
\end{eqnarray}
that is
\begin{equation}
  \sin{\varphi '\over 2}\sin\left(\alpha '-{\theta'\over 2}\right)=
\cos\chi\sin\alpha \sin{\psi\over 2}\,.\label{eq39}
\end{equation}
From Eqs. (\ref{eq37}) and (\ref{eq39}), since $\chi\ne \pm\pi /2$ and $\psi\ne 0\mod 2\pi$, we obtain
\begin{equation}
  \tan\left( \alpha '-{\theta'\over 2}\right)=\tan\alpha\,.\label{eq41}\end{equation}
and then
\begin{equation}
  \alpha '-{\theta'\over 2}=\alpha \mod \pi\,.\label{eq42}\end{equation}

If we choose $\alpha '-(\theta '/2)=\alpha$, then from Eq.\ (\ref{eq34}) we obtain
\begin{equation}
  \sin{\varphi ' \over 2}=\cos\chi\sin{\psi\over 2}\,,\label{eq43}\end{equation}
whereas if we choose  $\alpha '-(\theta '/2)=\alpha +\pi$, we obtain
\begin{equation}
  \sin{\varphi ' \over 2}=-\cos\chi\sin{\psi\over 2}\,.\label{eq44}\end{equation}
Since $\varphi '$ belongs to $[-\pi ,\pi ]$, changing $\alpha '=\alpha +(\theta' /2)$ into  $\alpha '=\alpha +(\theta' /2)+\pi$ changes $\varphi '$ into $-\varphi '$ and ${\q e}_m$ into $-{\q e}_m$, so that $\exp ({\q e}_m\varphi '/2)$ is unchanged. Solving Eq.\ (\ref{eq43}) or Eq.\ (\ref{eq44}) leads to the same result for $\varphi'$. Finally, we have to solve only Eqs.\ (\ref{eq15}) and (\ref{eq16}).

For the reciprocal formulas, we remark that Eq.\ (\ref{eq33}) is Eq.\ (\ref{eq27}),  and that choosing $\alpha '-(\theta '/2)=\alpha$, as above, leads to Eq.\ (\ref{eq29}). Eventually, Eqs.\ (\ref{eq34}) and (\ref{eq43}) lead to Eq.\ (\ref{eq28}).


\end{document}